\documentclass[aps,prd,twocolumn,superscriptaddress,amsfont,amssymb,amsmath,nofootinbib,showpacs,balancelastpage]{revtex4-1}
\usepackage{float}
\usepackage{epsfig}
\usepackage{graphicx}
\usepackage{dcolumn}
\usepackage{bm}
\usepackage{color}
\usepackage{tabularx}
\usepackage{multirow} 
\usepackage{booktabs} 
\usepackage{tabularx} 
\usepackage{cleveref}
\usepackage{hyperref,xcolor}
\hypersetup{
  colorlinks,
  citecolor=blue,
  linkcolor=blue,
  urlcolor=black}
\usepackage{xcolor}

\newcommand{\Od}{{\mathcal O}}
\newcommand{\Tr}{{\text{Tr}}}

\def\thebiblio#1{
\begin{center}\bf \large References
\end{center}
\list
{[\arabic{enumi}]}{\settowidth\labelwidth{#1.}\leftmargin\labelwidth
 \advance\leftmargin\labelsep
 \usecounter{enumi}}
 \def\newblock{\hskip .11em plus .33em minus -.07em}
 \sloppy
 \sfcode`\.=1000\relax}

\begin{document}

\preprint{}
\title{Gravitational perturbations of the Higgs field}

\author{Franco D.\ Albareti}
\email{`la Caixa'-Severo Ochoa Scholar, franco.albareti@csic.es}
\affiliation{Instituto de F{\'i}sica Te{\'o}rica UAM/CSIC, Universidad Aut{\'o}noma de Madrid, Cantoblanco, E-28049 Madrid, Spain}
\affiliation{Campus of International Excellence UAM+CSIC, Cantoblanco, E-28049 Madrid, Spain}

\author{Antonio L.\ Maroto}
\email{maroto@ucm.es}
\affiliation{Departamento de F\'{\i}sica Te\'orica, Universidad Complutense de Madrid, 28040 
Madrid, Spain}

\author{Francisco Prada}
\email{f.prada@csic.es}
\affiliation{Instituto de F{\'i}sica Te{\'o}rica UAM/CSIC, Universidad Aut{\'o}noma de Madrid, Cantoblanco, E-28049 Madrid, Spain}
\affiliation{Campus of International Excellence UAM+CSIC, Cantoblanco, E-28049 Madrid, Spain}
\affiliation{Instituto de Astrof{\'i}sica de Andaluc{\'i}a (CSIC), Glorieta de la Astronom{\'i}a, E-18080 Granada, Spain}

\date{\today}


\begin{abstract}

We study the possible effects of classical gravitational backgrounds on
the Higgs field through the modifications induced in the 
one-loop effective potential and the vacuum expectation value of the energy-momentum tensor.
We concentrate our study on the Higgs self-interaction contribution in a perturbed FRW metric. For weak and slowly varying gravitational
fields,  
a complete set of mode solutions for the Klein-Gordon equation is obtained to leading order in the 
adiabatic approximation. Dimensional regularization has been used in 
the integral evaluation and a detailed study of the integration of  
nonrational functions in this formalism has been presented. As expected, the regularized effective potential contains the same divergences as in 
flat spacetime, which can be renormalized without the need of  additional counterterms. We find that, in contrast with other regularization methods,  even though metric perturbations affect the  mode solutions,  they do not contribute to the leading adiabatic order of the potential. We also obtain explicit expressions of the complete energy-momentum tensor for general nonminimal coupling in terms of the perturbed modes. The corresponding leading adiabatic contributions are also obtained.

\end{abstract}


\maketitle


\section{Introduction}


There are two equally fundamental aspects of the Higgs mechanism 
for  electroweak symmetry breaking 
which have received remarkably different attention
in the last years.
On one hand, we have the prediction that  a new scalar boson should be present in the spectrum of the theory. Such
 a new particle has been  recently 
discovered by the ATLAS and CMS experiments at the LHC \cite{Higgs1,Higgs2}. The most
precise measurement to date of its mass comes from a combination of 
data from both experiments and is given by $m_H=125.09 \pm 0.21 (\mbox {stat})\pm 0.11 (\mbox{syst})$ GeV \cite{Hmass}. A large deal of experimental effort is being devoted to the study of the 
properties of the Higgs boson.  Apart from improving the precision in the 
determination of its 
mass, measurements of its production and decay channels, self-coupling and couplings to other particles are being performed. So far, all of them
are in excellent agreement with the predictions of the Standard Model (SM) \cite{Higgsprop,future,LHCP}.

On the other hand, the mechanism also predicts the existence of a Higgs field,
i.e.\ a constant classical field 
$\hat \phi=v$ with $v$ the Higgs vacuum expectation value (VEV)\footnote{In the SM, the 
Higgs VEV is  related to the 
Fermi coupling constant by $v= (\sqrt{2}\,G_F)^{-1/2}$. The value
of this constant is known since the original works of Fermi in the early  30's.} 
given by $v= 246.221 \pm 0.002$ GeV   
\cite{Donoghue}. It is precisely the interaction with the Higgs field what generates the 
masses of quarks, leptons and gauge bosons.  
The presence of this nonvanishing field which permeates all
of space is a  distinctive feature with respect to the
rest of SM fields  which have 
vanishing VEVs. Moreover, together with 
the homogeneous gravitational field created by the cosmological 
energy density, the Higgs field is the only SM field 
which is present today in the Universe on its largest scales. 
This fact opens the interesting possibility of probing the Higgs field
not only by exciting its quanta in colliders, but by directly perturbing its VEV. 
Thus, for example, the fact that the Higgs field 
 is a dynamical field sourced by massive particles 
 suggests that the presence of a heavy particle 
could induce shifts in  the
masses of  neighboring ones \cite{Heavy}. This effect
does not need the production of on-shell Higgs particles,
but because of the short range of the corresponding 
 Yukawa  interaction,  it is negligible at  distances beyond  the Compton wavelength of the Higgs boson. Existing data does not seem to contain enough kinematic information
in order to confirm or exclude it. A similar approach 
has been proposed in \cite{atoforce} in order to probe the 
Higgs couplings to electrons and light quarks.
The idea of generating peculiar
Higgs shifts was also considered in a different context in \cite{Onofrio}.
In that work a nonminimal coupling of the Higgs field 
to the spacetime curvature was considered. The nonminimal coupling 
modifies the effective potential inducing  shifts of the VEV 
in high-curvature regions such as those near neutron stars or black holes 
\cite{neutron}.

In this work, we explore further the effects of classical 
gravitational fields on the Higgs VEV. We consider
the SM Higgs minimally coupled to gravity. The Higgs VEV corresponds to the constant field configuration that  
minimizes the effective potential. This potential contains
not only the classical (tree-level)  contribution, but also loop corrections
introduced by quantum effects of all the particles that 
couple to the Higgs, including the Higgs self-interactions \cite{ColemanWeinberg}. More
relevant from the point of view of the present paper is the fact that these quantum corrections are sensitive to the spacetime geometry. The aim of this work is precisely to start the study of the Higgs one-loop effective
potential in weak and slowly varying gravitational
backgrounds. For simplicity and as a first step, we limit ourselves to the contributions of  the Higgs self-interactions. The fact that we assume  weak gravitational
backgrounds, i.e.\ whose curvature scale is much smaller than the Higgs mass, allows us to use an adiabatic approximation and 
avoid the problems generated by mode mixing and particle production
typical of quantum field theory in curved spacetime. For the same reason, we can still define an effective quasi-potential \cite{quasi,Hu} instead of using the full effective action since all the kinetic terms are suppressed with respect to the potential ones.  

Our work deals with the calculation of vacuum expectation values 
of quadratic operators in curved spacetime \cite{Birrell,Parker}. These are divergent
objects whose renormalization  requires the introduction of 
additional counterterms depending on the curvature tensors. Different
techniques have been used in the literature to work out these divergences
which, because of the fact that they are determined by the short-distance physics, depend locally on the geometry of spacetime \cite{SdW,SdW2,SdW3,SdW4,SdW5,ParkerFulling,ParkerFulling2,Ringwald,RGE,RGE2}.  But, apart from
the local divergent contributions, there are also finite 
nonlocal terms which are sensitive to the large-scale
structure of the manifold and, in general, depend on the 
quantum state on which the expectation value is evaluated. 
In some particular simple geometries, such as conformally flat 
metrics, these finite contributions can be exactly computed in some cases from 
the knowledge of the trace anomaly, but in general only brute force
methods, such as mode summation, are available to evaluate them
\cite{HuangBianchi,Huanginho,Huanginho2,Schwarzschild}. 
This is precisely the approach we follow in this work. 
In particular, we extend the analysis performed
in \cite{Maroto} to arbitrary dimension in order to calculate the integrals over the quantum modes using 
dimensional regularization. Several errors in \cite{Maroto} are also corrected in the present 
paper. 

The work is organized as follows: in Sec.\ \ref{back}, the effective action formalism is briefly reviewed. The field quantization in arbitrary $D+1$ dimensions in the adiabatic approximation is discussed in Sec.\ \ref{quan}. Section \ref{pert} contains the full mode solutions to first order in metric perturbations. The general results for the Higgs effective potential and the method used to obtain them are described in Sec.\ \ref{higgseff}. The vacuum expectation value of the energy-momentum tensor is calculated in Sec.\ \ref{tensor}. The paper ends in Sec.\ \ref{conc} with some discussions and conclusions.
 \vspace{-0.2cm}

\section{One-loop effective action}
\label{back}
The classical action for a minimally coupled real scalar field with potential $V(\phi)$ in
arbitrary \mbox{$(D+1)$}-dimensional curved spacetime reads
\begin{eqnarray}
S[\phi,g_{\mu\nu}]\,=\,\int \text{d}^{D+1}x \,\sqrt{g} \left(\frac{1}{2}\,g^{\mu\nu}\,\partial_\mu\phi\,\partial_\nu\phi\,-\,V(\phi)\right).\ \ \ 
\end{eqnarray}
In the case of the real Higgs field, the classical potential is given by
\begin{eqnarray}
V(\phi)\,=\,V_0\,+\,\frac{1}{2}M^2\phi^2\,+\,\frac{\lambda}{4}\phi^4
\label{potential}
\end{eqnarray}
with $M^2<0$. The minimum corresponds to  $\phi=v$ with $v^2=-M^2/\lambda$. The mass of the Higgs boson at tree-level   is given by $m_H^2=V''(v)=-2M^2$ and  from the recently measured value of $m_H$ at the LHC, the Higgs self-coupling
is  $\lambda\simeq 1/8$. 

The action is minimized by the solutions $\phi=\hat \phi$ of the classical equation of motion
\begin{eqnarray}
\Box \, \hat\phi \,+\,V'(\hat\phi)\, =\, 0 \,.
\label{KGc}
\end{eqnarray}
The quantum fluctuations around the classical solution
$\delta \phi=\phi-\hat\phi$ satisfy the equation of motion
\begin{eqnarray}
\left(\Box \, +\,m^2(\hat\phi)\right)\delta\phi\, =\, 0
\label{pKGc}
\end{eqnarray}
with
\begin{eqnarray}
m^2(\hat\phi)\,=\,V''(\hat\phi)\,=\,M^2\,+\,3\lambda \hat\phi^2\, .
\end{eqnarray}
The effective action which takes into account the effect 
of quantum fluctuations on the dynamics of the classical
field can be written as
\begin{eqnarray}
W[\hat\phi,g_{\mu\nu}]\,=\,\int \text{d}^{D+1}x \,\sqrt{g}\, L_{\text{eff}}
\end{eqnarray}
which can be expanded up to one-loop order as
\begin{eqnarray}
W[\hat\phi,g_{\mu\nu}]\,=\,S[\hat\phi,g_{\mu\nu}]\,+\,W^{(1)}[\hat\phi,g_{\mu\nu}]\,.
\end{eqnarray}
The one-loop correction $W^{(1)}$ can be written as
\cite{SdW5}
\begin{eqnarray}
W^{(1)}[\hat\phi,g_{\mu\nu}]\,=\,\frac{i}{2}\ln \det (-K)\,=\,\frac{i}{2}\Tr\ln (-K)
\end{eqnarray}
where $\Tr$ denotes the functional trace and $K$ is the quadratic operator associated to the quantum fluctuations
\begin{eqnarray}
K(x,y)\,=\,\left(\Box_x\,+\,m^2(\hat\phi)\right)\frac{\delta^{D+1}(x,y)}{\sqrt{g}}\,.
\end{eqnarray}

The corresponding Feynman's Green function
\begin{eqnarray}
i\,G_F(x,y)\,=\,\langle 0\vert T\left(\delta\phi(x)\delta\phi(y)\right)\vert 0\rangle
\end{eqnarray}
satisfies
\begin{eqnarray}
K(x,y)\,G_F(y,z)\,=\,-\frac{\delta^{D+1}(x,z)}{\sqrt{g}}
\label{GF}
\end{eqnarray}
where the de Witt repeated indices rule has been assumed.

Following \cite{Novikov,Zuk}, let us consider the derivative
of the one-loop effective action with respect to the 
mass parameter $m^2$, so that from (\ref{GF})
we can write 
\begin{eqnarray}
\frac{\text{d}W^{(1)}}{\text{d}m^2}\,=\,-\frac{i}{2}\Tr\,  G_F
\end{eqnarray}
or writing the trace explicitly
\begin{eqnarray}
\frac{\text{d}W^{(1)}}{\text{d}m^2}&=&-\frac{1}{2}\int \text{d}^{D+1}x\,\sqrt{g}\, i\, G_F(x,x)\nonumber \\
&=&-\frac{1}{2} \int \text{d}^{D+1}x\,\sqrt{g}\,\langle 0\vert \delta\phi^2(x)\vert 0\rangle\,.
\end{eqnarray}
Thus, we can finally get a formal expression for the one-loop contribution
to the effective Lagrangian as
\begin{eqnarray}
L_{\text{eff}}^{(1)}(x)\,=\,-\frac{1}{2}\int_0^{m^2(\hat\phi)} \text{d}m^2 \,\langle 0\vert\delta\phi^2(x)\vert 0 \rangle\,.
\label{Leff}
\end{eqnarray}
In general, in a static homogeneous spacetime, $\hat \phi$ is a constant field and the effective Lagrangian
defines the effective potential $V_1(\hat\phi)=-L_{\text{eff}}^{(1)}(\hat\phi)$.
In time-dependent or inhomogeneous spacetimes,
$\hat\phi$ changes in time or space and the effective potential is ill defined. In this case,  the effective
Lagrangian is a function of the 
classical fields; i.e., it will, in general, 
depend on $\hat\phi$ and $g_{\mu\nu}$  and arbitrary order
derivatives,
\begin{eqnarray}
L_{\text{eff}}\,=\,L_{\text{eff}}[\hat\phi,g_{\mu\nu},\partial\hat\phi,\partial^2\hat\phi,\partial g_{\mu\nu},\partial^2 g_{\mu\nu},\dots]\,.
\end{eqnarray}

However, in the case in which the background fields 
($\hat\phi$, $g_{\mu\nu}$) evolve very slowly in space and time compared to the evolution of the fluctuations,
the derivative terms in the effective Lagrangian are 
negligible, and the effective Lagrangian can be 
considered as an effective quasipotential \cite{quasi,Hu}. As we will
explicitly show in the next section, this is indeed the 
case for Higgs fluctuations in weak gravitational
backgrounds so that we can still 
define the one-loop effective potential as  
\begin{eqnarray}
V_{\text{eff}}(\hat\phi)\,=\,V(\hat\phi)\,+\,V_1(\hat\phi)
\end{eqnarray}
where
\begin{eqnarray}
V_1(\hat\phi)\,=\,-L_{\text{eff}}^{(1)}(\hat\phi)\,=\,\frac{1}{2}\int_0^{m^2(\hat\phi)} \text{d}m^2 \,\langle 0\vert\delta\phi^2\vert 0 \rangle\,.
\label{effpot}
\end{eqnarray}
The equation of motion for the classical 
field, thus, reduces to
\begin{eqnarray}
V'_{\text{eff}}(\hat\phi)\,\simeq\,0\,;
\label{V0}
\end{eqnarray}
i.e.\ the effective (quasi)potential correctly determines 
the VEV for a slowly varying background metric.

The central object in this calculation is the vacuum expectation value of a quadratic operator (\ref{Leff}). The standard Schwinger-de Witt representation
\cite{Birrell,Parker} allows us to obtain a local
expansion of $G_F$ in curvatures over the mass
parameter $m^2$. However, as mentioned before, this representation does not provide the full nonlocal
finite contributions of the effective action
in which we are interested in this work. Thus 
we will follow \cite{Birrell} and evaluate 
the expectation value from the explicit mode expansion
of the quantum fields.


\section{Quantization and adiabatic approximation}
\label{quan}
We will consider quantum fluctuations of the Higgs field 
in a \mbox{$(D+1)$}-dimensional spacetime metric which can be written as a scalar perturbation around  
 a flat Robertson-Walker background
\begin{eqnarray}
\text{d}s^2 \,=\,a^2(\eta) \left\{ \left[1 + 2 \Phi(\eta,{\bf x})\right]\, \text{d}\eta^2 - \left[1 - 2\Psi(\eta,{\bf x})\right]\,\text{d}{\bf x}^2 \right\}\nonumber \\\label{metric}
\end{eqnarray} 
where $\eta$ is the conformal time, $a(\eta)$ the scale factor, and $\Phi$ and $\Psi$ are the scalar perturbations in the longitudinal gauge.
This metric describes the spacetime geometry in cosmological contexts with density perturbations,  
but also, in the $a(\eta)=1$ case, it provides a good description 
of weak gravitational fields generated by slowly rotating  astrophysical objects like the Sun.

Up to first order in metric perturbations,  Eq.\ (\ref{pKGc}) for the fluctuation field $\delta\phi$ reads
\begin{eqnarray}
&\delta \phi''&+\left[(D-1)\,{\cal H}-\Phi'-D\,\Psi'\right]\delta\phi'
- \left[1+2(\Phi+\Psi)\right]\nabla^2\delta\phi \nonumber \\
&-&\boldsymbol{\nabla}\delta\phi\cdot\boldsymbol{\nabla}\left[\Phi-(D-2)\Psi\right] +a^2(1+2\Phi)\,m^2(\hat\phi)\,\delta\phi\,=\,0\,,\nonumber\\ \label{delta}
\end{eqnarray}
where ${\cal H}={a'}/a$ is the comoving Hubble parameter. 

 In order to  evaluate $V_1(\hat\phi)$, we need to quantize the 
fluctuation field. Because of the inhomogeneities of the 
metric tensor, exact solutions
for the perturbed Eq.\ \eqref{delta} are not expected to be found. 
Nevertheless, a perturbative expansion of the solution in powers of metric perturbations can be obtained. Moreover, 
when the  mode frequency $\omega$ is larger than the typical temporal or spatial frequency of the background metric, i.e.\ $\omega^2\gg {\cal H}^2$
and $\omega^2\gg \{\nabla^2\Phi,\nabla^2\Psi\}$, one can consider an adiabatic approximation in order to quantize the field fluctuations $\delta\phi$. Since
$\omega\geq m_H$,  
the adiabatic approximation is  extremely good  during the whole 
matter and acceleration eras until present, 
and also during most of the radiation era, 
for all  cosmological and astrophysical scales of interest. 

Let us start with the canonical quantization procedure for the field perturbations $\delta \phi$. Thus, following  \cite{Albareti,Albareti2}, 
we build a complete set of  mode solutions for \eqref{delta}, 
which are orthonormal  
with respect to the standard scalar product in curved spacetime 
\cite{Birrell}
\begin{eqnarray}
(\delta\phi_k,\delta\phi_{k'})\,=\,
i\int_\Sigma \left[ \delta\phi_{k'}^{*}\left(\partial_\mu \delta\phi_k\right)-\left(\partial_\mu \delta\phi_{k'}^{*}\right)\delta\phi_k\,
\right]\sqrt{g_\Sigma}\,
\text{d}\Sigma^\mu ,\nonumber \\
\label{scalar}
\end{eqnarray}
with $\text{d}\Sigma^\mu=n^\mu \text{d}\Sigma$. Here $n^\mu$ is a unit timelike vector
directed to the future and orthogonal to the $\eta=\text{const}$ hypersurface $\Sigma$, i.e.,
\begin{eqnarray}
\text{d}\Sigma^\mu\,=\,\text{d}^{D}{\bf x} \left(\frac{1-\Phi}{a},\bf{0}\right)\,, 
\end{eqnarray}
whereas the determinant of the metric on the 
spatial hypersurface reads to first order in metric perturbations
\begin{eqnarray}
\sqrt{g_\Sigma}\,=\,a^{D}(1-D\,\Psi)\,.
\end{eqnarray}
With this definition, the scalar product is independent on 
the choice of spatial hypersurface $\Sigma$. 

In terms of orthonormal modes,
\begin{eqnarray}
(\delta\phi_k,\delta\phi_{k'})\,=\,\delta^{D}({\bf k}-{\bf k'})\,,
\label{normalization}
\end{eqnarray}
the fluctuation field $\delta\phi$ can be expanded as
\begin{eqnarray}
\delta\phi(\eta,{\bf x})\,=\,\int \text{d}^{D}{\bf k} \left[ a_{{\bf k}}\,\delta\phi_k(\eta,{\bf x})+a^{\dag}_{{\bf k}}\,\delta\phi_k^*(\eta,{\bf x})\right].\ \ \ 
\end{eqnarray}
The corresponding creation and annihilation operators satisfy the
standard commutation relations
\begin{eqnarray}
[a_{{\bf k}},a^{\dag}_{{\bf k'}}]\,=\,\delta^{D}({\bf k}-{\bf k'})\,
\end{eqnarray}
and the vacuum state associated to the quantum modes $\{\delta\phi_k\}$ 
is defined as usual by $a_{\bf k}\vert 0\rangle=0$ $\forall {\bf k}$.

In order to construct the orthonormal set, we use a WKB ansatz,
\begin{eqnarray}
\delta\phi_k(\eta,{\bf x})\,=\,f_k(\eta,{\bf x}) \,e^{i\theta_k(\eta,{\bf x})}\,,
\label{wkb}
\end{eqnarray}
and assume that $f_k(\eta,{\bf x})$ evolves slowly 
in space and time, whereas the evolution of 
$\theta_k(\eta,{\bf x})$ is rapid. In general, as mentioned above, such an adiabatic ansatz  works whenever the Compton wavelength of the field 
perturbation is much smaller than the typical astrophysical or cosmological scales involved.
In particular, in the adiabatic expansion we assume 
$\partial \theta\sim ma$ and  $\partial f\sim {\cal H}f$.

Substituting \eqref{wkb} in \eqref{delta}, we obtain to the leading adiabatic order $\Od((\partial \theta)^2)$
\begin{eqnarray}
-\theta'^2_k+\left[1+2(\Phi+\Psi)\right](\boldsymbol{\nabla}\theta_k)^2+m^2\,a^2(1+2\Phi)\,=\,0\,\ \ \ \ \ \  \label{leading}
\end{eqnarray}
and to the next-to-leading order $\Od(\partial\theta)$
\begin{eqnarray}
f_k\theta''_k &+& 2f'_k\theta'_k +\left[(D-1)\,{\cal H}-\Phi'-D\,\Psi'\right] f_k\theta'_k\nonumber \\
&-&f_k\nabla^2\theta_k - 2\boldsymbol{\nabla} f_k\cdot\boldsymbol{\nabla}\theta_k\, \\
&-&f_k \,\boldsymbol{\nabla}\theta_k\cdot \boldsymbol{\nabla}\left[\Phi-(D-2)\Psi\right]\nonumber
=\,0\,. \label{next}
\end{eqnarray}
Notice that $\partial^2\theta\sim {\cal H}\,\partial \theta$ and
that, in the adiabatic expansion, ${\cal H}\sim \partial\Phi$.


\section{Perturbative expansion and mode solutions}
\label{pert}

To solve these two equations, \eqref{leading} and \eqref{next}, we 
look for a perturbative expansion in the metric potentials.
To obtain the lowest-order solution; i.e., in the 
absence of metric perturbations, we write \eqref{delta} in the
limit $\Phi=\Psi=0$ and get
\begin{eqnarray}
\delta {\phi^{(0)}}''+(D-1)&\,{\cal H}&\,\delta{\phi^{(0)}}'- \nabla^2\delta\phi^{(0)}
+a^2 m^2(\hat\phi)\,\delta\phi^{(0)}=0 ,\nonumber\\
\end{eqnarray}
where $a^2m^2(\hat\phi)$ only depends on time. Fourier transforming the spatial coordinates, the following positive frequency solution with momentum ${\bf k}$ is obtained
\begin{eqnarray}
\delta\phi^{(0)}_k(\eta,{\bf x})\,=\,F_k(\eta)\, e^{i{\bf k}\cdot{\bf x}-i\int^\eta\omega_k(\eta')\text{d}\eta'} 
\end{eqnarray}
with
\begin{eqnarray}
\omega_k^2\,&=&\,k^2\,+\,m^2\,a^2 \label{omega} 
\end{eqnarray}
and 
\begin{eqnarray}
F_k(\eta)\,&=&\,\frac{1}{(2\pi)^{D/2}}\,\frac{1}{a^{(D-1)/2}\,\sqrt{2\,\omega_k}}\,\label{Fsol}
\end{eqnarray}
which is fixed by the normalization condition \eqref{normalization}.

Once the unperturbed solution is known, we can look for the 
first-order corrections. Thus, the amplitude and phase of \eqref{wkb}  are expanded in metric perturbations as follows
\begin{eqnarray}
f_k(\eta,{\bf x})&\,=\,&F_k(\eta)\,+\,\delta f_k(\eta,{\bf x})\nonumber \\ \\
\theta_k(\eta,{\bf x})&\,=\,&{\bf k}\cdot {\bf x}\,-\int^\eta\omega_k(\eta')\,\text{d}\eta'\,+\,\delta\theta_k(\eta,{\bf x})\,\nonumber \label{at}
\end{eqnarray}
where $\delta f_k$ and $\delta\theta_k$ are first order in 
perturbations. 
Substituting \eqref{at} in the leading equation \eqref{leading}, we obtain  \eqref{omega} to the lowest order as expected, 
and to first order we get
\begin{eqnarray}
\omega_k\,\delta\theta'_k+{\bf k}\cdot \boldsymbol{\nabla}\delta\theta_k
+k^2(\Phi+\Psi)+m^2\,a^2\,\Phi\,=\,0\,.\label{leadingpert}
\end{eqnarray}
On the other hand, by substituting in the next-to-leading 
equation \eqref{next}, we recover \eqref{Fsol} to the lowest perturbative order, whereas to first order  we get
\begin{eqnarray}
&&F_k\, \delta\theta''_k+ 2F'_k\,\delta\theta'_k +(D-1)\,{\cal H}\,F_k \,\delta\theta'_k -F_k\,\nabla^2\delta\theta_k\nonumber \\ 
&-& 2\,\omega_k\,\delta f'_k-2\,{\bf k}\cdot \boldsymbol{\nabla}\delta f_k
-(D-1)\,\omega_k\,{\cal H}\,\delta f_k -\omega_k'\delta f_k \label{nonleadingpert} \\ 
&+& \omega_k\, F_k\,\Phi'+D\,\omega_k \,F_k\,\Psi'
- F_k \,{\bf k}\cdot \boldsymbol{\nabla}\left[\Phi-(D-2)\,\Psi\right]\,=\,
0\,.\nonumber 
\end{eqnarray}

The two new equations \eqref{leadingpert}
and \eqref{nonleadingpert} can also be solved by performing an additional Fourier transformation in the 
spatial coordinates since the equations  
coefficients only depend on time. 


\subsection*{Phase solution $\delta\theta_k$}

Equation \eqref{leadingpert} in Fourier space reads
\begin{eqnarray}
\delta\theta'_k(\eta,{\bf p})+i\frac{{\bf k}\cdot{\bf p}}{\omega_k}\delta\theta_k(\eta,{\bf p})\,=\,-\omega_k\left[\Phi(\eta,{\bf p})+\frac{k^2}{\omega_k^2}\Psi(\eta,{\bf p})\right],\nonumber\\
\label{deltagen}
\end{eqnarray}
where 
\begin{eqnarray}
\delta\theta_k(\eta,{\bf p})\,=\,\frac{1}{(2\pi)^{3/2}}\int \text{d}^3{\bf x} \,
\delta\theta_k(\eta,{\bf x})\,e^{-i{\bf p}\cdot{\bf x}}\, 
\end{eqnarray}
and analogous definitions apply for $\Phi(\eta,{\bf p})$, $\Psi(\eta,{\bf p})$ and $\delta f_k(\eta,{\bf p})$.\footnote{In the following, the wave vector of the 
 quantum modes is denoted by ${\bf k}$, and ${\bf p}$ is used for that of metric perturbations.}
Defining
\begin{eqnarray}
\beta_k(\eta_f,\eta_i)&\,=\,&\int_{\eta_i}^{\eta_f} \frac{\text{d}\eta'}{\omega_k(\eta')}  \nonumber \\ \\
G_k(\eta,{\bf p})&\,=\,&-\omega_k\left[\Phi(\eta,{\bf p})\,+\,\frac{k^2}{\omega_k^2}\,\Psi(\eta,{\bf p})\right], \nonumber
\end{eqnarray}
the solution of \eqref{deltagen} is
%
%
\begin{eqnarray}
\label{dtheta}
&&\delta\theta_k(\eta,{\bf p})\,=\,\\ &&\int_{0}^\eta e^{-i\,{\bf k}\cdot {\bf p}\,\beta_k(\eta,\eta')} \, G_k(\eta',{\bf p})
\,\text{d}\eta'+e^{-i\,{\bf k}\cdot {\bf p}\,\beta_k(\eta,0)}\delta\theta_k(0,{\bf p})\,.\nonumber
\end{eqnarray}
The term $\delta\theta_k(0,{\bf p})$ stands for the initial boundary condition of the modes or, equivalently, the phase difference of the modes at the initial time. In principle, $\delta\theta_k(0,{\bf p})$ is not completely arbitrary since the orthonormalization condition of the modes \eqref{normalization} may constrain its functional dependence. We discuss this point at the end of this section.


\subsection*{Amplitude solution $\delta f_k$}

Let us write
\begin{eqnarray}
\delta f_k(\eta,{\bf p})\,=\,F_k(\eta)\,P_k(\eta,{\bf p})\,
\label{df}
\end{eqnarray}
and following a similar procedure with the next-to-leading order
equation \eqref{nonleadingpert}, it can be rewritten in Fourier space as
\begin{eqnarray}
P'_k(\eta,{\bf p})\,+\,i\,\frac{{\bf k}\cdot{\bf p}}{\omega_k}\,P_k(\eta,{\bf p}) \,=\,\frac{H_k(\eta,{\bf p})}{2\,\omega_k}\,,
\end{eqnarray}
where
\begin{eqnarray}
H_k(\eta,{\bf p})\,=\,\omega_k\, Q'_k(\eta,{\bf p})\,+\,T_k(\eta,{\bf p})
\end{eqnarray}
with
\begin{eqnarray}
Q_k(\eta,{\bf p})\,=\,-i\frac{{\bf k}\cdot{\bf p}}{\omega_k^2}\,\delta\theta_k(\eta,{\bf p})+ 
\left[D-\frac{k^2}{\omega_k^2}\right]\Psi(\eta,{\bf p})\ \ \ 
\end{eqnarray}
and 
\begin{eqnarray}
T_k(\eta,{\bf p})&\,=\,&p^2\,\delta\theta_k(\eta,{\bf p})\\
&-&i\,{\bf k}\cdot {\bf p}\left[\Phi(\eta,{\bf p})-(D-2)\,\Psi(\eta,{\bf p})\right]\nonumber\,.
\end{eqnarray}
%
The corresponding solution is given by
\begin{eqnarray}
\label{Pk}
&&P_k(\eta,{\bf p})\,=\,\\
&&\int_0^\eta e^{-i\,{\bf k}\cdot{\bf p}\,\beta_k(\eta,\eta')}\frac{H_k(\eta',{\bf p})}{2\omega_k(\eta')}\,\text{d}\eta'+e^{-i\,{\bf k}\cdot{\bf p}\,\beta_k(\eta,0)} P_k(0,{\bf p})\,.\nonumber
\end{eqnarray}
The integration constant $P_k(0,{\bf p})$ is fixed by the normalization
condition \eqref{normalization}.


\subsubsection*{Time-independent gravitational potentials}

For simplicity, in the rest of the work we focus on time-independent gravitational potentials. This case encompasses 
super-Hubble modes
in both matter and radiation era, and also sub-Hubble modes in the 
matter era. This is also a good approximation to describe
the gravitational potentials in the Solar System.
In such a case, the constants $P_k(0,{\bf p})$ are given by
\begin{eqnarray}
P_k(0,{\bf p})\,=\, \frac{1}{2}\left(D-\frac{k^2}{\omega_k(0)^2}\right)\Psi({\bf p})\,.
\end{eqnarray}
%


 Integrating by parts in \eqref{Pk}, the integration
 constant can be eliminated and  the following expression
 is obtained:
\begin{eqnarray}
&&P_k(\eta,{\bf p})\,=\,
\frac{1}{2}Q_k(\eta,{\bf p})-\\&&i\int_0^\eta\left\lbrace\frac{{\bf k}\cdot{\bf p}}{2\,\omega_k(\eta')}
 \,e^{-i\,{\bf k}\cdot{\bf p}\,\beta_k(\eta,\eta')}
  \left[Q_k(\eta',{\bf p}) 
  + \frac{T_k(\eta',{\bf p})}{{\bf k}\cdot{\bf p}}\right]\right\rbrace\text{d}\eta'\,.\nonumber
 \end{eqnarray}

There are three types of contributions to $P_k$, depending
on the number of time integrals involved. Thus, we can write
\begin{eqnarray}
P_k(\eta,{\bf p})\,=\,P_k^{(0)}(\eta,{\bf p})+P_k^{(1)}(\eta,{\bf p})+P_k^{(2)}(\eta,{\bf p})
 \end{eqnarray}
where  
\begin{eqnarray}
P_k^{(0)}(\eta,{\bf p})\,&=&\,\frac{1}{2}\left(D-\frac{k^2}{\omega_k(\eta)^2}\right)\Psi({\bf p})\label{p0}\\
\ \nonumber\\
P_k^{(1)}(\eta,{\bf p})\,&=&\,
\int_0^\eta e^{-i\,{\bf k}\cdot{\bf p}\,\beta_k(\eta,\eta')}\,N_k^{(1)}(\eta,\eta',{\bf p})
\,\text{d}\eta' \label{p1}\\
P_k^{(2)}(\eta,{\bf p}) \,&=&\,
\int_0^\eta\int_0^{\eta'} e^{-i\,{\bf k}\cdot{\bf p}\,\beta_k(\eta,\eta'')}\,N_k^{(2)}(\eta',\eta'',{\bf p})\,
\text{d}\eta'' \,\text{d}\eta'\nonumber \\\label{p2}
 \end{eqnarray} 
 with
  \begin{eqnarray}
&&N_k^{(1)}(\eta,\eta',{\bf p})\,=\,\frac{i{\bf k}\cdot{\bf p}}{2\,\omega_k^2(\eta)\,\omega_k(\eta')}\times\\&&\left\lbrace\left[\omega_k^2(\eta')-\omega_k^2(\eta)\right]\Phi({\bf p})\vphantom{\left[k^2+\omega_k^2(\eta)\left(\frac{k^2}{\omega_k^2(\eta')}-2\right)\right]}
\right.\nonumber
 +\left.\left[k^2+\omega_k^2(\eta)\left(\frac{k^2}{\omega_k^2(\eta')}-2\right)\right]\Psi({\bf p})\right\rbrace
\end{eqnarray}
\begin{eqnarray}
N_k^{(2)}(\eta',\eta'',{\bf p})\,=\,\frac{({\bf k}\cdot{\bf p})^2-p^2\omega_k^2(\eta')}{2\,\omega_k^3(\eta')\,\omega_k(\eta'')}&&\times\ \ \ \ \ \ \ \ \ \ \ \\\left[\omega_k^2(\eta'')\,\right.&&\left.\Phi({\bf p})+k^2\,\Psi({\bf p})\right]\nonumber
 \end{eqnarray} 
where $p=|{\bf p}|$.
 %
%


\subsection*{Orthonormalization condition}


In order to quantize the field canonically, we must check that the modes $\delta\phi_k$ used to define the creation and annihilation operators are orthonormal \eqref{normalization}. This may restrict the functional dependence of the initial conditions of our solution, i.e., $P_k(0,{\bf p})$ and $\delta\theta_k(0,{\bf p})$\footnote{$P_k(0,{\bf x})$ and $\delta\theta_k(0,{\bf x})$ are assumed to be real. If this were not the case, the phase of $P_k(0,{\bf x})$ could be absorbed into $\delta\theta_k(0,{\bf x})$ and the imaginary part of $\delta\theta_k(0,{\bf x})$ could also be absorbed into $P_k(0,{\bf x})$ in a trivial way.}. We already fixed $P_k(0,{\bf p})$ when imposing the correct normalization of the modes; hence, we can only play with $\delta\theta_k(0,{\bf p})$ to have orthogonal modes.
The scalar product \eqref{scalar} can be computed using \eqref{wkb}, \eqref{at}, \eqref{dtheta}, and \eqref{Pk} to get 
\begin{eqnarray}
(\delta\phi_k,\delta\phi_{k'})\,=\,\delta^{D}({\bf k}-{\bf k'})\,+\,\tau_\Psi({\bf k},{\bf k'})\,+\,\tau_{\delta\theta}({\bf k},{\bf k'})\,,\ \ \ \ 
\label{tau}
\end{eqnarray}
where $\tau_{\Psi,\delta\theta}$ are first order in metric perturbation. The explicit expressions for $\tau_{\Psi,\delta\theta}$ are given in Appendix \ref{app0}. In this appendix it is shown that they are zero for $\forall {\bf k},{\bf k'}$ up to corrections beyond the leading adiabatic order for slowly varying gravitational fields. This result does not impose any restriction on the functional dependence of $\delta\theta_k(0,{\bf p})$.

Different initial conditions $\delta\theta_k(0,{\bf p})$ amount to different definitions of the vacuum. The discussion above guarantees that the modes given by \eqref{wkb}, \eqref{at}, \eqref{dtheta}, \eqref{Pk}, are orthonormal for any choice of the vacuum. In the following we take $\delta\theta_k(0,{\bf p})=0$ as the initial condition for the modes.

\section{Higgs effective potential}
\label{higgseff}

Once we have the expressions for the mode solutions of the 
perturbative equations, namely  \eqref{dtheta} and \eqref{df} [together with \eqref{Pk}];  we can proceed to calculate the one-loop contribution
to the effective potential \eqref{effpot}.

Let us first calculate $\langle 0\vert\delta\phi^2(\eta, {\bf x})\vert 0 \rangle$ to first order in metric perturbations. Because of the inhomogeneity 
of the background, this quantity depends on $(\eta, {\bf x})$ as follows
\begin{eqnarray}
&&\langle 0\vert\delta\phi^2(\eta, {\bf x})\vert 0 \rangle\,=\,\langle\delta\phi^2\rangle_{\text{h}}(\eta)\,+\,\langle\delta\phi^2\rangle_{\text{i}}(\eta, {\bf x})
\end{eqnarray}
where
\begin{eqnarray}
&&\langle\delta\phi^2\rangle_{\text{h}}(\eta)\,=\,\int \text{d}^{D}{\bf k}\,F_k^2(\eta)
\end{eqnarray}
and
\begin{eqnarray}
\langle\delta\phi^2\rangle_{\text{i}}(\eta, {\bf x})\,=\,
 2\int \text{d}^{D}{\bf k}\, F^2_k(\eta)\,\left[ \text{Re}P_k(\eta, {\bf x})-\text{Im}\,\delta\theta_k(\eta, {\bf x})\right].\nonumber\\
 \label{deltai1}
\end{eqnarray}
%


\subsubsection*{Homogeneous contribution $\langle\delta\phi^2\rangle_{\rm{h}}$}

The homogeneous contribution $\langle\delta\phi^2\rangle_{\text{h}}$ reads
\begin{eqnarray}
\langle\delta\phi^2\rangle_{\text{h}}(\eta)\,&=&\,\frac{1}{2\,(2\pi)^D\,a^{D-1}(\eta)}\int   \frac{\text{d}^D{\bf
k}}{\sqrt{k^2+m^2\,a^2(\eta)}}\\
\,&=&\,\frac{1}{2\,(2\pi)^D\,a^{D-1}(\eta)}\,\frac{2\,\pi^{D/2}}{\Gamma(D/2)}\int_0^\infty \frac{\text{d}k\,k^{D-1}}{\sqrt{k^2+m^2\,a^2(\eta)}}\nonumber\,
\label{deltah}
\end{eqnarray}
which is analogous to the Minkowskian result, except for the scale-factor dependence.


\subsubsection*{Nonhomogeneous contribution $\langle\delta\phi^2\rangle_{\rm{i}}$}

The inhomogeneous component $\langle\delta\phi^2\rangle_\text{i}$ can be dealt with more easily in momentum 
space. The only
angular dependence of the quantum fluctuation wave vector ${\bf k}$ enters as ${\bf k}\cdot{\bf p}=k\,p\,\hat x$ with $\hat x=\cos \theta$, where 
we have taken the $k_z$ direction along  ${\bf p}$. On the other hand, the contribution from $\delta\theta$  in (\ref{deltai1})
vanishes after integrating in $\hat x$. Then, we have
\begin{eqnarray}
\label{deltaiD}
\langle\delta\phi^2\rangle_{\text{i}}(\eta, {\bf p})\,&=&\,\frac{1}{(2\pi)^D\,a^{D-1}(\eta)}\int \text{d}^D{\bf k}\,  \frac{P_k(\eta,{\bf p})}{\sqrt{k^2+m^2\,a^2(\eta)}}\,. \nonumber \\
\end{eqnarray}
Since the integration on $\hat x$  can be performed in a straightforward way, let us define
\begin{eqnarray}
\hat{P}_k (\eta,{\bf p})\,=\,\int_{-1}^1 \text{d}\hat{x}\,\left(1-\hat{x}^2\right)^{(D-3)/2}\,P_k (\eta,{\bf p})\,
\end{eqnarray}
where we have included the general integration measure in $D$ dimensions.
Hence, we can write (see Appendix \ref{app1})
\begin{eqnarray}
\label{deltai}
\langle\delta\phi^2\rangle_{\text{i}}(\eta, {\bf p})=\frac{1}{(2\pi)^D\,a^{D-1}(\eta)}&&\,\frac{2\,\pi^{(D-1)/2}}{\Gamma((D-1)/2)}\\&&\times\int_0^\infty \text{d}k\,\frac{k^{D-1}\,\hat{P}_k(\eta,{\bf p})}{\sqrt{k^2+m^2\,a^2(\eta)}}. \nonumber
\end{eqnarray}

Both integrals \eqref{deltah} and \eqref{deltai} are divergent in $D=3$ dimensions and they should be regularized as discussed in the next section.  


\subsection*{Regularization}

Let us now discuss the regularization procedure based on 
standard dimensional regularization techniques.


\subsubsection*{Regularized homogeneous contribution $\langle\delta\phi^2\rangle_{\rm{h}}(\eta)$}

The momentum integral in $\langle\delta\phi^2\rangle_{\text{h}}$ \eqref{deltah} can be done using (\ref{redux}) of Appendix \ref{app1}. After expanding for small $\epsilon$ with $D=3-\epsilon$ dimensions, the final result is
\begin{eqnarray}
\langle\delta\phi^2\rangle_{\text{h}}(\eta)\,=\,
\frac{m^2(\hat\phi)}{16\pi^2}\left[\ln\left(\frac{m^2(\hat\phi)}{\mu^2}\right)-N_\epsilon-\frac{3}{2}\right]
\label{v1h}
\end{eqnarray}
where $\mu$ is the renormalization scale and
\begin{eqnarray}
N_\epsilon\,=\,\frac{2}{\epsilon}+\log 4\pi-\gamma
\end{eqnarray}
with $\gamma$ the Euler-Mascheroni constant.


\subsubsection*{Regularized nonhomogeneous contribution $\langle\delta\phi^2\rangle_{\rm{i}}(\eta,{\bf x})$}

Let us now consider the inhomogeneous contribution \eqref{deltai}. We cannot apply directly standard dimensional regularization formulas  because of the nontrivial $k$ dependence of $\hat P_k(\eta,{\bf p})$. Thus, additional work is necessary.

First, it should be noticed that the dependence of $\hat{P}_k(\eta,{\bf p})$ on the direction of ${\bf p}$ only enters through the potentials, $\Phi({\bf p})$ and $\Psi({\bf p})$.
Therefore, it can be expanded in the following way:
\begin{eqnarray}
\label{pls}
\,& \hat{P}_k (\eta,{\bf p})\, =\hspace{4cm}\\ &\left[\sum_{l=0}^{\infty}  P_{k,l}^\Phi(\eta)\,p^{2l}\right]\,\Phi({\bf p})\,+\, \left[\sum_{l=0}^{\infty}  P_{k,l}^\Psi(\eta)\,p^{2l}\right]&\,\Psi({\bf p})\nonumber\,.
\end{eqnarray}
The coefficients $P_{k,l}^{\left\lbrace\Phi,\Psi\right\rbrace}(\eta)$ are given in Appendix \ref{app2}. The $l=0$ terms only get
contributions from the $P_k^{(0)}(\eta,{\bf p})$ term given in \eqref{p0}, and its integral vanishes in dimensional regularization. The $l>0$
terms involve time integrals of the form
\begin{eqnarray}
\label{f1}
\int_0^\eta \text{d}\eta' \left(\prod_{i=1}^{2l-1}\int_{\eta'}^{\eta}\frac{\text{d}\eta_i}{\omega_k(\eta_i)} \right)\frac{k^{2\alpha}}{\omega_k(\eta)^a\,\omega_k(\eta')^b}\,
\end{eqnarray}
for the contributions coming from $P_k^{(1)}(\eta,{\bf p})$ in  \eqref{p1}, and
\begin{eqnarray}
\label{f2}
\int_0^\eta \text{d}\eta' \int_0^{\eta'} \text{d}\eta'' \left(\prod_{i=1}^{2l-2}\int_{\eta''}^{\eta} \frac{\text{d}\eta_i}{\omega_k(\eta_i)} \right)\frac{k^{2\alpha}}{\omega_k(\eta)^a\,\omega_k(\eta')^b\,\omega_k(\eta'')^c}
\nonumber \\
\end{eqnarray}
for those coming from $P_k^{(2)}(\eta,{\bf p})$ in \eqref{p2}, with $\alpha,a,b,c\in \mathbb{Z}$. 
In order to simplify the functional dependence on $k$, we apply the generalized Feynman trick,
\begin{eqnarray}
\frac{1}{A_1^{d_1}\cdots A_n^{d_n}}\,&=&\,\frac{\Gamma(d_1+\cdots+d_n)}{\Gamma(d_1)\cdots\Gamma(d_n)}\,\int_0^1\text{d}x_1\cdots\int_0^1\text{d}x_n\ \ \ \ \ \ \ \\
\times \, \delta\left(x_1+\cdots \right.&+&\left.x_n-1\right)
 \frac{x_1^{d_1-1}\cdots x_n^{d_n-1}}{\left(x_1\,A_1+\cdots+x_n\,A_n\right)^{d_1+\cdots+d_n}}.\nonumber
\end{eqnarray}
%
%
Then, let the parameters of the Feynman formula be defined by
\begin{eqnarray}
n\,&=&\,2l+1\\
A_j\,&=&\,
\left\{
	\begin{array}{ll}
		\omega_k^2(\eta)\ \ \ \ \ \ &\text{if}\ j=1\\
\omega_k^2(\eta')\ \ \ &\text{if}\ j=2\\
\omega_k^2(\eta_{j-2})\ \ \ &\text{if}\ 3\leq j \leq 2l+1
	\end{array}
\right.\\
d_j\,&=&\,
\left\{
	\begin{array}{ll}
		a/2\ \ \ \ \ \ \ \ \ \ &\text{if}\ j=1\\
		b/2\ \ \ &\text{if}\ j=2\\
1/2\ \ \ &\text{if}\ 3\leq j \leq 2l+1
	\end{array}
\right.
\end{eqnarray}
for the case \eqref{f1} [with a trivial modification for the expression \eqref{f2}]. In this way, the $k$ dependence only appears in $\sum_{i=1}^{2l+1} x_i\,\omega^2_{k,i}$ which can be simplified in the following way,
\begin{eqnarray}
\sum_{i=1}^{2l+1} x_i\,\omega^2_{k,i}\,=\,\sum_{i=1}^{2l+1} x_i\, (k^2+m^2\,a_i^2)\,=k^2+m^2\sum_{i=1}^{2l+1} x_i\,a_i^2\,,\ \ \ \ \
\end{eqnarray}
where we have used $\sum_{i=1}^{2l+1} x_i= 1$. Now, the $k$ dependence is simple enough to use standard dimensional regularization formulas (Appendix \ref{app1}). The integration over the $\{x_i\}$ and the time integrals can be performed analytically (Appendix \ref{app3}).

As we did with $\hat P_k(\eta,{\bf p})$, we now decompose $\langle\delta\phi^2\rangle_{\text{i}} (\eta,{\bf p})$ into two terms proportional to  $\Phi({\bf p})$ and $\Psi({\bf p})$, respectively,
\begin{eqnarray}
\langle\delta\phi^2\rangle_{\text{i}}(\eta,{\bf p})\,=\,\langle\delta\phi^2\rangle^\Phi_{\text{i}}(\eta,{\bf p})\,\Phi({\bf p})\,+\,\langle\delta\phi^2\rangle^\Psi_{\text{i}}(\eta,{\bf p})\,\Psi({\bf p})\,. \nonumber\\\label{Deltai}
\end{eqnarray}
%
Then, integrating in dimensional regularization, we see that the  ${\cal O}(1/\epsilon)$ terms cancel out, and the results are finite
\begin{eqnarray}
\langle\delta\phi^2\rangle^{\left\lbrace\Phi,\Psi\right\rbrace}_{\text{i}}(\eta,{\bf p})\,=\,\frac{m^2}{4\pi^2 a^2(\eta)}\,\left[\sum_{l=1}^{\infty}  R_{l}^{\left\lbrace\Phi,\Psi\right\rbrace}(\eta)\,p^{2l}\right],\ \ \ \ 
\label{regphi}
\end{eqnarray}
where $R_{l}^{\left\lbrace\Phi,\Psi\right\rbrace}$ are the already regularized integrals in $k$ of $P_{k,l}^{\left\lbrace\Phi,\Psi\right\rbrace}$ divided by $m^2$ for convenience. The coefficients $R_{l}^{\left\lbrace\Phi,\Psi\right\rbrace}$ can be written as
\begin{eqnarray}
R^{\left\lbrace\Phi,\Psi\right\rbrace}_{l}(\eta)\,=\,R^{\left\lbrace\Phi,\Psi\right\rbrace}_{l,\,\text{pol}}(\eta)\,+\,R^{\left\lbrace\Phi,\Psi\right\rbrace}_{l,\,\text{log}}(\eta)\,,
\label{rparts}
\end{eqnarray}
where, as shown in Appendix \ref{app3}, $R^{\left\lbrace\Phi,\Psi\right\rbrace}_{l,\,\text{pol}}$  are polynomials in $\eta$, and $R^{\left\lbrace\Phi,\Psi\right\rbrace}_{l,\,\text{log}}$ involve a logarithmic dependence on $\eta$. 

The most important aspect of \eqref{regphi} is that all the divergent parts  have canceled out. In particular, the divergent  terms coming from $P^{(1)}_k(\eta,{\bf p})$ cancel exactly the ones from $P^{(2)}_k(\eta,{\bf p})$ order by order in $p$. This means that the UV behavior is the same as in an unperturbed FRW background and the inhomogeneous contributions  are 
finite to the leading adiabatic order.


\subsection*{Nonhomogeneous contribution: Particular cases}
\label{cases}

\subsubsection*{Nonexpanding spacetimes}
Let us consider weak gravitational fields generated by  static 
sources. For the corresponding
spacetime metric, we can take 
(\ref{metric}) with $a(\eta)=1$ and static potentials $\Phi({\bf x})$ and $\Psi({\bf x})$ which allow us to  use the previous results. This simplifies the calculations in several of the steps discussed above. For instance, all the time integrals can be done in a straightforward way, there is no need to apply the Feynamn trick since the $\omega$'s are all the same, and the coefficients $R^{\left\lbrace\Phi,\Psi\right\rbrace}_{l,\text{log}}$ are zero (see Appendix \ref{app3}).

The results for a nonexpanding geometry read
\begin{eqnarray}
R^\Phi_l(\eta)\,&=&\,R^\Phi_{l,\text{pol}}(\eta)\,=0\,\,\\
R^\Psi_l(\eta)\,&=&\,R^\Psi_{l,\text{pol}}(\eta)\,=0\,\,,
\end{eqnarray}
which imply
\begin{eqnarray}
\langle\delta\phi^2\rangle^\Phi_{\text{i}}(\eta,{\bf p})\,&=&\,0 \ \ \\
\langle\delta\phi^2\rangle^\Psi_{\text{i}}(\eta,{\bf p})\,&=&\,0 \,
\end{eqnarray}
and
\begin{eqnarray}
\langle\delta\phi^2\rangle_{\text{i}}(\eta,{\bf p})\,&=&\,0\,.
\end{eqnarray}
Thus, to the leading adiabatic order, the metric perturbations do 
not contribute to the Higgs effective potential in dimensional regularization. This is in contrast with previous results \cite{Maroto}
using cutoff regularization, in which nonvanishing inhomogeneous 
contributions were obtained. 
 
Although we have considered a particular coordinate choice in
  (\ref{metric}), corresponding to the longitudinal gauge, 
  since in the absence of metric 
  perturbations $V_{\text{eff}}^{\text{h}}(\hat\phi)$ is a 
  constant, the Stewart-Walker lemma \cite{S-W} guarantees that the obtained
  effective potential is gauge invariant.

\subsubsection*{Expanding spacetimes: Cosmology}
Now we consider the case of a perturbed  expanding universe with 
scale factor $a(\eta)$ and constant metric perturbations 
$\Phi({\bf x})$ and $\Psi({\bf x})$. In particular, we will concentrate in  the matter-dominated era ,in which the metric perturbations are constant both for sub-Hubble and super-Hubble modes. In addition, we will also provide results for super-Hubble modes  in the radiation era for which the metric perturbations are also constant.

For the $\Psi$ contribution, we get for the matter and radiation eras
with $a\propto \eta^2$ and $a\propto \eta$, respectively,
\begin{eqnarray}
R^{\Psi}_{l,\text{pol}}(\eta)\,=\,0\;;\ \  R^{\Psi}_{l,\text{log}}(\eta)\,=\,0 \,.
\label{psicos}
\end{eqnarray}

The $\Phi$ terms are harder to compute since the $R^{\Phi}_{l,\,\text{log}}$ contribution is not zero, and the integration over the Feynman parameters $\{x_i\}$ and the time integrals has to be performed by Taylor expanding the logarithm (see Appendix \ref{app3}). An exact analytical expression can be obtained for each order of the logarithm expansion given in terms of finite sums, which can be computed numerically for practical purposes. We have checked that the relative difference between $R^\Phi_{l,\,\text{pol}}$ and $R^{\Phi}_{l,\,\text{log}}$ terms is $\sim 10^{-4}$ for $l=1,2,3$ and $\sim 10^{-2}$ for $l=4,5$. Then,
\begin{eqnarray}
\frac{R^{\Phi}_{l,\text{pol}}(\eta)\,+\,R^{\Phi}_{l,\text{log}}(\eta)}{R^{\Phi}_{l,\text{pol}}(\eta)}\,\leq \,10^{-2}\nonumber\,\hspace{.5cm}\text{for}\ l=1,2,3,4,5.
\end{eqnarray}
This suggests that the exact $\Phi$ contribution also may be zero as for the $\Psi$ terms, so that for expanding geometries as well, static perturbations do not contribute to the Higgs effective potential to the 
leading adiabatic order. 

\subsection*{Higgs effective potential}

Taking into account \eqref{effpot}, the one-loop contribution to the effective potential can be expressed as
\begin{eqnarray}
V_{1}(\eta,{\bf x})\,=\,
 V_1^{\text{h}}(\eta)+V_1^{\text{i}}(\eta, {\bf x})\,.
\end{eqnarray}

Given the fact that, to the leading order the nonhomogeneous contribution 
vanishes, the potential  reads
\begin{eqnarray}
V_1=V_1^{\text{h}}(\eta)\,=\,\frac{1}{2}\int_0^{m^2(\hat\phi)}\text{d}m^2\,\langle\delta\phi^2\rangle_{\text{h}}(\eta)\,,
\label{RWint}
\end{eqnarray}
and substituting (\ref{v1h}), we get

\begin{eqnarray}
V_1(\hat\phi)\,=\,
\frac{m^4(\hat\phi)}{64\pi^2}\left[\ln\left(\frac{m^2(\hat\phi)}{\mu^2}\right)-N_\epsilon-\frac{3}{2}\right].
\end{eqnarray}
As expected from previous works \cite{SdW,SdW2,SdW3,SdW4,SdW5,ParkerFulling,ParkerFulling2,Ringwald,RGE,RGE2}, the homogeneous contribution is constant even though the geometry is 
expanding.  The $N_\epsilon$  term is proportional to $m^4(\hat \phi)$, so that 
we have three kinds of divergences: constant,
quadratic in $\hat \phi$ and quartic, which can be 
reabsorbed in the renormalization of the tree-level potential
parameters  
$V_0$, $M^2$ and $\lambda$. This means that at the leading adiabatic
order we obtain exactly the same divergences as
in flat spacetime and  we do not need additional counterterms to renormalize
the effective potential.  

Following the minimal subtraction scheme $\overline{\text{MS}}$,
 we remove the terms proportional to $N_\epsilon$.
Thus, we are
left with the complete renormalized homogeneous effective potential,
\begin{eqnarray}
&&V_{\text{eff}}(\hat\phi)\,=\,\\&&V_0+\frac{1}{2}M^2\,\hat\phi^2+\frac{\lambda}{4}\,\hat\phi^4 \label{Vren}
+
\frac{m^4(\hat\phi)}{64\pi^2}\left[\ln\left(\frac{m^2(\hat\phi)}{\mu^2}\right)-\frac{3}{2}\right] ,\nonumber
\end{eqnarray}
which agrees with the standard result in flat spacetime.
Here, the physical mass $M$ and coupling constant $\lambda$ are defined
at a given physical scale $\mu$.
Since the renormalized effective potential is independent of the renormalization
scale $\mu$, $M^2$ and the coupling 
constant should depend on $\mu$ according to the renormalization group equations
\begin{eqnarray}
\beta(\lambda)&\equiv&\frac{\text{d}\lambda}{\text{d}(\log \mu)}\,=\,\frac{18\lambda^2}{(4\pi)^2}\nonumber\\\label{RGE} \\
\gamma_M(\lambda)&\equiv&\frac{\text{d}(\log M^2)}{\text{d}(\log \mu)}=\frac{6\lambda}{(4\pi)^2}\,.\nonumber
\end{eqnarray}

\section{Energy-momentum tensor}
\label{tensor}
In the previous sections, we have considered the one-loop correction 
to the effective potential. The complete set of perturbed modes obtained
also allows us to evaluate the  vacuum expectation value of the 
energy-momentum tensor. For the sake of completeness we will include also a 
possible nonminimal coupling to curvature, so that the 
equation for an arbitrary massive scalar field now reads 
\begin{eqnarray}
\left(\Box\,+\,m^2\,+\,\xi R\right)\varphi\,=\,0\,,
\end{eqnarray}
Notice that to the leading adiabatic order, the curvature term is not
going to modify the mode solutions found in Sec.\ IV; however,
 the energy-momentum tensor acquires new contributions. Thus,
\begin{widetext}
\begin{eqnarray}
T^\mu_{\;\nu}\,=\,-\delta^\mu_{\;\nu}\left(\frac{1}{2}\,-\,2\,\xi\right)\left(g^{\rho\sigma}\partial_\rho\varphi\,
\partial_\sigma\varphi-m^2\varphi^2\right)\,+\,\left(1-2\,\xi\right)\,g^{\mu\rho}\partial_\rho\varphi
\,\partial_\nu\varphi\,-2\,\xi\,\varphi\,\nabla^\mu\nabla_\nu\varphi\,+\\+\,\frac{2}{D+1}\xi\,g^\mu_{\;\nu}\left(\varphi\Box\varphi\,+\,m^2\varphi^2\right)\,-\,\xi\left(R^\mu_{\;\nu}\,-\,\frac{1}{2}R\,g^\mu_{\;\nu}\,+\,\frac{2D}{D+1}\xi\,R\,g^\mu_{\;\nu}\right)\varphi^2\,.\nonumber\ \ \ \ \ \ 
\end{eqnarray}
\end{widetext}
Considering perturbations over a flat Robertson-Walker background
$\eqref{metric}$, the vacuum expectation value of this tensor, $\langle T^\mu_{\;\;\nu} \rangle$, can be explicitly written to the leading adiabatic order in Fourier space as a mode sum in terms of the expansion \eqref{at}  as 
\begin{widetext}
\begin{eqnarray}
\langle T^0_{\;0} (\eta,{\bf p})\rangle\,&=&\,\rho(\eta,{\bf p})\,=\,\frac{1}{(2\pi)^D}\frac{1}{a^{D+1}}\int \text{d}^D {\bf k}\,\frac{\omega_k}{2}\left[1\,+\,2\,\frac{k^2}{\omega_k^2}\,\Psi({\bf p})\,+\,2\,P_{k}(\eta,{\bf p})\,+\,2\,i\,\frac{{\bf k\cdot p}}{\omega_k^2}\,\delta\theta_k(\eta,{\bf p})\,-\,\frac{2\,\xi}{\omega_k^2}\,P''_k(\eta,{\bf p})\right]\nonumber\\ \\\nonumber\\
\langle T^i_{\;i}(\eta,{\bf p})\rangle\,&=&\,-p_i(\eta,{\bf p})\,=\,-\frac{1}{(2\pi)^D}\frac{1}{a^{D+1}}\int \text{d}^D\, {\bf k}\,\left[\frac{k^2_i}{2\,\omega_k}\left(1\,+\,2\,\Psi({\bf p})\,+\,2\,P_{k}(\eta,{\bf p})\vphantom{\frac{1}{1}}\right)\,+\,2\,i\,\frac{k_i\,p_i}{2\,\omega_k}\,\delta\theta_k(\eta,{\bf p})\,+\,\xi\,\frac{p_i^2}{\omega_k}\,P_k(\eta,{\bf p})\right]\nonumber\\ \\\nonumber\\
\langle T^i_{\;0}(\eta,{\bf p})\rangle\,&=&\,\frac{1}{(2\pi)^D}\frac{1}{a^{D+1}}\int \text{d}^D {\bf k}
\left[\frac{k_i}{2}\left(1\,+\,2\,P_k(\eta,{\bf p})\,+\,2\,i\frac{{\bf k}\cdot{\bf p}}{2\,\omega_k^2}\delta\theta_k(\eta,{\bf p})\right)+
\frac{i}{2}\,p_i\,\delta\theta_k(\eta,{\bf p})\,+\,\xi\,\frac{i\,p_i}{\omega_k}\,P'_k(\eta,{\bf p})
\right]\\\nonumber\\
\langle T^i_{\;j}(\eta,{\bf p})\rangle\,&=&\,-\frac{1}{(2\pi)^D}\frac{1}{a^{D+1}}\int \text{d}^D {\bf k}\left[\frac{k_i\,k_j}{2\,\omega_k}\left(1\,+\,2\,\Psi({\bf p})\,+\,2\,P_{k}(\eta,{\bf p})\vphantom{\frac{1}{1}}\right)\,+\,i\,\frac{k_i\,p_j+k_j\,p_i}{2\,\omega_k}\,\delta\theta_k(\eta,{\bf p})\,+\,\xi\,\frac{p_i\,p_j}{\omega_k}\,P_k(\eta,{\bf p})\right]\\
\nonumber\\
\langle T^\mu_{\;\mu}(\eta,{\bf p})\rangle\,&=&\,\frac{1}{(2\pi)^D}\frac{1}{a^{D+1}}\int \text{d}^D {\bf k}\,\left[\frac{m^2}{2\,\omega_k}\left(1\,+\,2\,P_k(\eta,{\bf p})\vphantom{\frac{1}{1}}\right)\,-\,\frac{\xi}{\omega_k}\left(P''_k(\eta,{\bf p})\,+\,p^2\,P_k(\eta,{\bf p})\right)\right]
\end{eqnarray}

\end{widetext}

The integration over the quantum modes can be performed using the same methods applied above and some tricks to reduce the integrals involving the components of ${\bf k}$, $k_i$ or $k_i\,k_j$, to integrals of scalar character in ${\bf k}$ (Appendix \ref{app1}). After doing that,
the homogeneous part is found to be diagonal, and the energy $\rho$ and pressure $p$ are given in the minimal substraction scheme $\overline{\text{MS}}$ by
\begin{eqnarray}
\rho\,=-\,p\,=\,\frac{m^4}{64\,\pi^2}\,\left[\log\left(\frac{m^2}{\mu^2}\right)\,-\,\frac{3}{2}\right],
\end{eqnarray}
where $\mu$ is the renormalization physical scale.

On the other hand, much as for the effective potential, the nonhomogeneous part of the energy-momentum tensor vanishes to this order. 

While classical and weak gravitational fields are not able to change the UV behavior of quantum effects, it is expected that gravity should  modify the IR parts of all quantum corrections.  
The result presented in this work shows that, within the dimensional regularization scheme, there are no gravitational corrections arising from a perturbed FRW metric up to first order in perturbations, and  to the leading  order in the adiabatic expansion, to the vacuum expectation value of the energy-momentum tensor of a scalar field. Then, gravitational corrections may appear beyond the leading adiabatic order, or through nonlinear terms.

In the considered regime, namely the one in which the Hubble scale is much smaller than the mass of the quantum field, corrections beyond the zero adiabatic order are negligible and they are unlikely to belong to the experimental realm in the near future.

On the other hand, although nonlinear contributions are expected to be smaller than the linear ones, they will be more important than the contribution from the first adiabatic order. Nevertheless, the computation of the second-order corrections to the energy-momentum tensor is a formidable task which is well beyond the scope of this work.


\section{Discussion and conclusions}
\label{conc}

In this work, we have computed the one-loop corrections to the effective potential due to the self-interactions of the Higgs field and the 
vacuum expectation value of its energy-momentum tensor in a perturbed FRW background. Unlike previous results 
based on the Schwinger-de Witt approximation, we have calculated 
explicitly a complete orthonormal set of  modes of the perturbed Klein-Gordon equation and the dimensional regularization procedure was used
for the mode summation to the leading adiabatic order.  The   integrals containing metric perturbations involved nonrational functions of the momenta so that standard formulas in dimensional regularization were not suitable to evaluate them.  New  expressions have been developed for those cases which applied both to static and expanding backgrounds.

We have checked that 
the homogeneous contribution agrees with the Minkowski result as expected.
On the other hand, we have found that to the leading adiabatic order, and 
to first order in metric perturbations, no additional contributions appear
either in the regularized effective potential nor in the energy-momentum tensor. This is in contrast with previous results obtained with a cutoff regularization \cite{Maroto}, in which quartic and quadratic inhomogeneous divergences 
appear in the calculation. Thus, we see that dimensional regularization ensures that the theory can be renormalized just absorbing the 
divergences in the tree-level parameters (at the leading adiabatic order).

We expect additional contributions from the metric perturbations at the next-to-leading adiabatic orders. Unlike the Schwinger-de Witt method which provides a local expansion of the effective action. The mode summation method used in this work could allow to determine the corresponding finite nonlocal 
contributions. In this sense, the explicit mode calculation obtained here
together with the method developed to perform the integrals in dimensional 
regularization of nonrational functions of the momenta  are a fundamental first step in this program. The results presented in this work would also 
allow to calculate the temperature effects on the Higgs effective potential 
 using the explicit mode summation and, in general,  
the complete expressions of other expectations values in perturbed metric
backgrounds. Work is in progress in these directions.

\ \\

{\it Acknowledgements}. This work has been supported by the Spanish MICINNs Consolider-Ingenio 2010 Programme under grant MultiDark CSD2009-00064, MINECO Centro de Excelencia Severo Ochoa Programme under grant SEV-2012-0249, and MINECO grants FIS2014-52837-P,  AYA-2012-31101 and AYA2014-60641-C2-1-P. FDA acknowledges financial support from `la Caixa'-Severo Ochoa doctoral fellowship.
\appendix
\onecolumngrid


\section{Orthonormalization condition: $\tau_\Psi({\bf k}, {\bf k'})$ and $\tau_{\delta\theta}({\bf k}, {\bf k'})$}
\label{app0}
In this appendix, we show that $\tau_\Psi({\bf k}, {\bf k'})$ and $\tau_{\delta\theta}({\bf k}, {\bf k'})$ appearing in \eqref{tau} are zero to the leading adiabatic order. This implies that the modes given by \eqref{wkb}, \eqref{at}, \eqref{dtheta}, \eqref{df}, \eqref{Pk}, are orthonormal and, therefore, the scalar field $\delta\phi$ can be quantized within the canonical formalism.

The explicit expressions for $\tau_\Psi({\bf k}, {\bf k'})$ and $\tau_{\delta\theta}({\bf k}, {\bf k'})$ are
\begin{eqnarray}
\tau_\Psi({\bf k},{\bf k'})\,&=&\,\int \text{d}^D{\bf x}\,\frac{\left(\omega_k-\omega_{k'}\right)\left(k^2\,\omega_{k'}^2\,-\,k'^2\,\omega_k^2\right)}{4\,(\omega_k\,\omega_{k'})^{5/2}}\,\Psi({\bf x})\,\frac{e^{i\,({\bf k}-{\bf k'})\cdot {\bf x}}}{(2\pi)^D}\\\nonumber \\
\tau_{\delta\theta}({\bf k},{\bf k'})\,&=&\,\int \text{d}^D{\bf x}\,\frac{\left(\omega_k-\omega_{k'}\right)\,\left(\omega_{k'}^2\,{\bf k}\cdot\nabla\delta\theta_k(0,{\bf x})\,-\,\omega_k^2\,{\bf k'}\cdot\nabla\delta\theta_{k'}(0,{\bf x})\right)}{4\,(\omega_k\,\omega_{k'})^{5/2}}\,\frac{e^{i\,({\bf k}-{\bf k'})\cdot {\bf x}}}{(2\pi)^D}
\end{eqnarray}
First, let us focus on $\tau_\Psi$ in Fourier space:
\begin{eqnarray}
\tau_\Psi({\bf k},{\bf k'})\,&=&\,\int \text{d}^D{\bf x}\int \frac{\text{d}^D{\bf p}}{(2\pi)^{D/2}}\,\frac{\left(\omega_k-\omega_{k'}\right)\left(k^2\,\omega_{k'}^2\,-\,k'^2\,\omega_k^2\right)}{4\,(\omega_k\,\omega_{k'})^{5/2}}\,\Psi({\bf p})\,\frac{e^{i\,({\bf k}-{\bf k'}+{\bf p})\cdot {\bf x}}}{(2\pi)^D}\nonumber\\
\nonumber\\
\,&=&\,\int \frac{\text{d}^D{\bf p}}{(2\pi)^{D/2}}\,\frac{\left(\omega_k-\omega_{k'}\right)\left(k^2\,\omega_{k'}^2\,-\,k'^2\,\omega_k^2\right)}{4\,(\omega_k\,\omega_{k'})^{5/2}}\,\Psi({\bf p})\,\delta^D({\bf k}-{\bf k'}+{\bf p})\nonumber\\
\nonumber\\
\,&=&\,\frac{1}{(2\pi)^{D/2}}\,\frac{\left(\omega_k-\omega_{k'}\right)\left(k^2\,\omega_{k'}^2\,-\,k'^2\,\omega_k^2\right)}{4\,(\omega_k\,\omega_{k'})^{5/2}}\,\Psi({\bf k}-{\bf k'})
\end{eqnarray}
Since $\Psi$ varies over macroscopic scales, we can a assume an exponential damping for $\Psi$ when $|{\bf k}-{\bf k'}|\gg|\nabla\Psi|\sim{\cal H}$; therefore, $\tau_\Psi({\bf k},{\bf k'})\,\approx\,0$ in this case. For $|{\bf k}-{\bf k'}|\sim{\cal H}$, we can Taylor expand the coefficient in front of $\Psi({\bf k}-{\bf k'})$ in ${\cal H}/\omega_k$ to get
\begin{eqnarray}
\tau_\Psi({\bf k},{\bf k'})\,&\approx&\,\frac{1}{(2\pi)^{D/2}}\frac{m^2\,k^2}{2\,\omega^4_k}\left(\frac{{\cal H}}{\omega_k}\right)^2\Psi({\bf k}-{\bf k'})
\end{eqnarray}
which is beyond the leading adiabatic order.

The same procedure works for $\tau_{\delta\theta}$, for instance,
\begin{eqnarray}
\tau_{\delta\theta}({\bf k},{\bf k'})\,&=&\,\int \text{d}^D{\bf x}\int \frac{\text{d}^D{\bf p}}{(2\pi)^{D/2}}\,\frac{\left(\omega_k-\omega_{k'}\right)\,i\,{\bf p}\cdot\left(\omega_{k'}^2\,{\bf k}\,\delta\theta_k(0,{\bf p})\,-\,\omega_k^2\,{\bf k'}\,\delta\theta_{k'}(0,{\bf p})\right)}{4\,(\omega_k\,\omega_{k'})^{5/2}}\,\frac{e^{i\,({\bf k}-{\bf k'}+{\bf p})\cdot {\bf x}}}{(2\pi)^D}\nonumber\\
\nonumber\\
\,&=&\,\int \frac{\text{d}^D{\bf p}}{(2\pi)^{D/2}}\,\frac{\left(\omega_k-\omega_{k'}\right)\,i\,{\bf p}\cdot\left(\omega_{k'}^2\,{\bf k}\,\delta\theta_k(0,{\bf p})\,-\,\omega_k^2\,{\bf k'}\,\delta\theta_{k'}(0,{\bf p})\right)}{4\,(\omega_k\,\omega_{k'})^{5/2}}\,\delta^D({\bf k}-{\bf k'}+{\bf p})\nonumber\\
\nonumber\\
\,&=&\,\frac{1}{(2\pi)^{D/2}}\,\frac{\left(\omega_k-\omega_{k'}\right)\,i\,\left({\bf k}-{\bf k'}\right)\cdot\left(\omega_{k'}^2\,{\bf k}\,\delta\theta_k(0,{\bf k}-{\bf k'})\,-\,\omega_k^2\,{\bf k'}\,\delta\theta_{k'}(0,{\bf k}-{\bf k'})\right)}{4\,(\omega_k\,\omega_{k'})^{5/2}}
\end{eqnarray}
The initial condition is supposed to not introduce power at small scales; therefore, $\delta\theta_{k}(0,{\bf k}-{\bf k'})$ is also exponentially damped for modes $|{\bf k}-{\bf k'}|\gg{\cal H}$. For $|{\bf k}-{\bf k'}|\sim{\cal H}$, we can Taylor expand in ${\cal H}/\omega_k$ to get
\begin{eqnarray}
\tau_{\delta\theta}({\bf k},{\bf k'})\,\approx\,\frac{i}{(2\pi)^{D/2}}\,\frac{1}{4\,\omega_k^3}\,{\bf k}\cdot\frac{{\bf k}-{\bf k'}}{|{\bf k}-{\bf k'}|}\left(\frac{{\cal H}}{\omega_k}\right)^3\,\left(\left(m^2-k^2\right)\delta\theta_k(0,{\bf k}-{\bf k'})+\omega_k^2\,{\bf k}\cdot \nabla \delta\theta_k(0,{\bf k}-{\bf k'})\vphantom{\frac{1}{1}}\right)\,.
\label{tautheta}
\end{eqnarray}
Thus, for $|{\bf k}-{\bf k'}|\sim{\cal H}$, $\tau_{\delta\theta}$ is also beyond the leading adiabatic order. Note that the nabla operator in \eqref{tautheta} is to be understood as acting over the index variable $k$, not over the argument ${\bf k}-{\bf k'}$.


\section{Dimensional Regularization Formulas}
\label{app1}

The fundamental formula used in dimensional regularization in Euclidean space is \cite{DR,DR1}
\begin{eqnarray}
\int\frac{\text{d}^D{\bf k}}{(2\pi)^D}\,\frac{k^{2\alpha}}{\left(k^2+m^2\right)^\beta}\,=\,m^{2(\alpha-\beta)}\,\left(\frac{m^2}{4\pi}\right)^{D/2}\,\frac{\Gamma(D/2+\alpha)\,\Gamma(\beta-\alpha-D/2)}{\Gamma(\beta)\,\Gamma(D/2)}\,.
\end{eqnarray}
This expression has been used to compute $\langle\delta\phi^2\rangle_{\text{h}}$ in  \eqref{deltah} in $D=3-\epsilon$. The left-hand side of the equation can be written as
\begin{eqnarray}
\int\frac{\text{d}^D{\bf k}}{(2\pi)^D}\,\frac{k^{2\alpha}}{\left(k^2+m^2\right)^\beta}\,=\,\frac{1}{(2\pi)^D}\,\frac{2\pi^{D/2}}{\Gamma(D/2)}\,\int_0^\infty\text{d}k\,\frac{k^{D-1}\,k^{2\alpha}}{(k^2+m^2)^{\beta}}
\end{eqnarray}
then
\begin{eqnarray}
\int_0^\infty\text{d}k\,\frac{k^{D-1}\,k^{2\alpha}}{(k^2+m^2)^{\beta}}\,=\,\left[\frac{1}{(2\pi)^D}\,\frac{2\pi^{D/2}}{\Gamma(D/2)}\right]^{-1}\,m^{2(\alpha-\beta)}\,\left(\frac{m^2}{4\pi}\right)^{D/2}\,\frac{\Gamma(D/2+\alpha)\,\Gamma(\beta-\alpha-D/2)}{\Gamma(\beta)\,\Gamma(D/2)}\,\,.
\label{redux}
\end{eqnarray}

On the other hand, for the $\langle\delta\phi^2\rangle_{\text{i}}$ term in  \eqref{deltaiD}, we have to deal with integrals of the following form
\begin{eqnarray}
\int\frac{\text{d}^D{\bf k}}{(2\pi)^D}\,\frac{f({\bf k}\cdot{\bf p})}{\left(k^2+m^2\right)^\beta}
\label{cos}
\end{eqnarray}
where $f({\bf k}\cdot {\bf p})$ is an analytical function. Taking the $k_z$ direction along ${\bf p}$, we have $f({\bf k}\cdot{\bf p})=f(k\,p\,\hat x)$ with $k=|{\bf{k}}|$, $p=|{\bf{p}}|$ and $\hat x=\cos(\theta_{D-2})$, $\theta_{D-2}$ being the angle between ${\bf k}$ and ${\bf p}$. When using spherical coordinates in $D$ dimensions $\left\lbrace\phi,\theta,\theta_2,...,\theta_{D-2}\right\rbrace$, the volume element can be expressed as
\begin{eqnarray}
\text{d}^D{\bf k}\,=\,k^{D-1}\,\sin^{D-2}(\theta_{D-2})\,\sin^{D-3}(\theta_{D-3})...\sin(\theta)\,\text{d}k\,\text{d}\phi\,\text{d}\theta...\,\text{d}\theta_{D-2}.
\end{eqnarray}
The integrand of \eqref{cos} depends on $\cos(\theta_{D-2})$, so we can integrate in all the angular variables but $\theta_{D-2}$. With that purpose, notice that the area of a sphere in a $D$-dimensional space is
\begin{eqnarray}
\overbrace{\int_0^\pi...\int_0^\pi}^{D-2}\int_0^{2\pi}\sin^{D-2}(\theta_{D-2})\,\sin^{D-3}(\theta_{D-3})...\sin^2(\theta_2)\,\sin(\theta)\,\text{d}\phi\,\text{d}\theta\,\text{d}\theta_2...\,\text{d}\theta_{D-2}\,=\frac{2\pi^{D/2}}{\Gamma(D/2)}.
\end{eqnarray}
Since all the integrals involved can be factorized, the integration over all the angular variables but $\theta_{D-2}$ is simply given by the area of a sphere in $(D-1)$-dimensional space, i.e.\ $\frac{2\pi^{(D-1)/2}}{\Gamma((D-1)/2)}$.
%
%
Therefore, Eq.\ \eqref{cos} can be expressed as
\begin{eqnarray}
\int\frac{\text{d}^D{\bf k}}{(2\pi)^D}\,\frac{f({\bf k}\cdot{\bf p})}{\left(k^2+m^2\right)^\beta}\,=\,\frac{1}{(2\pi)^D}\,\frac{2\,\pi^{(D-1)/2}}{\Gamma((D-1)/2)}\,\int_{0}^{\infty}\text{d}k\,\frac{k^{D-1}}{(k^2+m^2)^\beta}\,\hat{f}(k\,p)\,, \label{kp}
\end{eqnarray}
where $\hat{f}(k\,p)=\int_{-1}^{1}\text{d}\hat x\,\left(1-\hat{x}^2\right)^{(D-3)/2}\,f(k\,p\,\hat{x})$. Finally, Taylor expanding $\hat{f}(k\,p)$, the expression can be regularized order by order using Eq.\ \eqref{redux}.

To regularize physical quantities like $\langle\delta\phi^2\rangle_{\text{h}}$ and $\langle\delta\phi^2\rangle_{\text{i}}$ two important aspects should be taken into consideration. First of all, the full physical expression should be computed in $D$ dimensions, so that when taking $D=3-\epsilon$, all the terms are expanded in $\epsilon$. Moreover, a physical scale $\mu^{\epsilon}$ should be introduced to compensate the physical dimensions. 

\subsection*{Integrals involving $k_i$ or $k_i k_j$}

Finally, we explain how to compute the integrals involving the components of ${\bf k}$, $k_i$ and $k_i\,k_j$, appearing in the expression of the energy-momentum tensor in Sec.\ \ref{tensor}. For these cases, the other vector quantity, namely the wave vector of the metric perturbations ${\bf p}$, can be used to produce scalar quantities that can be easily computed in terms of the expressions given above. For instance
\begin{eqnarray}
\int\text{d}^D{\bf k}\,g(k,{\bf k}\cdot{\bf p})\,k_i \,=\,A\,p_i\,,
\end{eqnarray}
taking the scalar product with ${\bf p}$ in each member we get that
\begin{eqnarray}
A\,=\,\int\text{d}^D{\bf k}\,g(k,{\bf k}\cdot{\bf p})\,\frac{{\bf k}\cdot{\bf p}}{p^2}
\end{eqnarray}
which can be integrated using the expression \eqref{kp}. For the remaining case, we have
\begin{eqnarray}
\int\text{d}^D{\bf k}\,g(k,{\bf k}\cdot{\bf p})\,k_i\,k_j =\,B\,\delta_{ij}\,+\,C\,p_i\,p_j\,,
\end{eqnarray}
where $B$ and $C$ can be computed solving the system obtained by taking the trace and contracting with $p^i\,p^j$. The results are
\begin{eqnarray}
B\,&=&\,\frac{1}{(D-1)}\int\text{d}^D{\bf k}\,g(k,{\bf k}\cdot{\bf p})\,\frac{\left(k\,p\right)^2-\left({\bf k}\cdot{\bf p}\right)^2}{p^2}\\
C\,&=&\,\frac{1}{(D-1)}\int\text{d}^D{\bf k}\,g(k,{\bf k}\cdot{\bf p})\,\frac{D\left({\bf k}\cdot{\bf p}\right)^2-\left(k\,p\right)^2}{p^4}\,.
\end{eqnarray}
%


\section{$P^{\left\lbrace\Phi,\,\Psi\right\rbrace}_{k,l}$}
\label{app2}

In this appendix, the exact expressions for the $P_{k,\,l}^{\left\lbrace \Phi,\,\Psi\right\rbrace}(\eta)$ coefficients of Eq.\ \eqref{pls} are given. First, let us separate these coefficients as
\begin{eqnarray}
P^\Phi_{k,l}(\eta)\,=\,P^{{\Phi},(0)}_{k,l}(\eta)+P^{\Phi,(1)}_{k,l}(\eta)+P^{\Phi,(2)}_{k,l}(\eta)\,,
\end{eqnarray}
where the indices $(0), (1), (2)$ stand for the contribution coming from $P^{(0)}_k$ in \eqref{p0}, $P^{(1)}_k$ in \eqref{p1}, and $P^{(2)}_k$ in \eqref{p2}, respectively. The same definition applies for the terms $P^{\Psi}_{k,l}$.

The $l=0$ coefficients are given by
\begin{eqnarray}
P^{\Phi}_{k,\,0}(\eta)\,&=&\,0\\
P^{\Psi}_{k,\,0}(\eta)\,&=&\,P^{\Psi,\,(0)}_{k,\,0}(\eta)\,=\,\frac{1}{2}\,\frac{\sqrt{\pi}\,\Gamma((D-1)/2)}{\Gamma(D/2)}\left(D-\frac{k^2}{\omega_k^2(\eta)}\right).
\end{eqnarray}
For $l>0$, we have
\begin{eqnarray}
P^{\Phi}_{k,\,l}(\eta)\,&=&\,P^{\Phi,\,(1)}_{k,\,l}(\eta)\,+\,P^{\Phi,\,(2)}_{k,\,l}(\eta)\,\\
P^{\Psi}_{k,\,l}(\eta)\,&=&\,P^{\Psi,\,(1)}_{k,\,l}(\eta)\,+\,P^{\Psi,\,(2)}_{k,\,l}(\eta)\,\label{ppsi}
\end{eqnarray}
with
\begin{eqnarray}
P^{\Phi,\,(1)}_{k,\,l}(\eta)\,&=&\,\frac{(-1)^{l}}{2^{2l}}\,\frac{\sqrt{\pi}\,\Gamma((D-1)/2)}{(l-1)!\,\Gamma(D/2+l)}\,k^{2l}\,\int_0^\eta\text{d}\eta'\left(\prod_{i=1}^{2l-1}\int_{\eta'}^{\eta}\frac{\text{d}\eta_i}{\omega_k(\eta_i)}\right)\left[\frac{1}{\omega_k(\eta')}-\frac{\omega_k(\eta')}{\omega_k^2(\eta)}\right]\,\\
P^{\Psi,\,(1)}_{k,\,l}(\eta)\,&=&\,\frac{(-1)^{l}}{2^{2l}}\,\frac{\sqrt{\pi}\,\Gamma((D-1)/2)}{(l-1)!\,\Gamma(D/2+l)}\,k^{2l}\,\int_0^\eta\text{d}\eta'\left(\prod_{i=1}^{2l-1}\int_{\eta'}^{\eta}\frac{\text{d}\eta_i}{\omega_k(\eta_i)}\right)\left[\frac{2}{\omega_k(\eta')}-\frac{k^2}{\omega_k^2(\eta)\,\omega_k(\eta')}-\frac{k^2}{\omega_k^3(\eta')}\right]\,\label{ppsi1}\\
P^{\Phi,\,(2)}_{k,\,l}(\eta)\,&=&\,\frac{(-1)^{l}}{2^{2l-1}}\,\frac{\sqrt{\pi}\,\Gamma((D-1)/2)}{(l-1)!\,\Gamma(D/2+l-1)}\,k^{2l}\,\int_0^\eta\text{d}\eta'\int_0^{\eta'}\text{d}\eta''\left(\prod_{i=1}^{2l-2}\int_{\eta''}^{\eta}\frac{\text{d}\eta_i}{\omega_k(\eta_i)}\right)\left[\frac{\omega_k(\eta'')}{k^2\,\omega_k(\eta')}-\frac{(2l-1)\,\omega_k(\eta'')}{(D-2l-2)\,\omega_k^3(\eta')}\right]\,\\
P^{\Psi,\,(2)}_{k,\,l}(\eta)\,&=&\,\frac{(-1)^{l}}{2^{2l-1}}\,\frac{\sqrt{\pi}\,\Gamma((D-1)/2)}{(l-1)!\,\Gamma(D/2+l-1)}\,k^{2l}\,\int_0^\eta\text{d}\eta'\int_0^{\eta'}\text{d}\eta''\left(\prod_{i=1}^{2l-2}\int_{\eta''}^{\eta}\frac{\text{d}\eta_i}{\omega_k(\eta_i)}\right)\left[\frac{1}{\omega_k(\eta')\,\omega_k(\eta'')}-\frac{(2l-1)\,k^2}{(D-2l-2)\,\omega_k^3(\eta')\,\omega_k(\eta'')}\right].\nonumber\\\label{ppsi2}
\end{eqnarray}

The integral over $k$ of all these terms can be regularized with  the expressions given in Appendix \ref{app1} after applying the generalized Feynman trick discussed in Sec.\ \ref{higgseff}. After regularization, we are left with two terms: one polynomic in $\eta$, the other one logarithmic in $\eta$. The integration over the Feynman parameters $\{x_i\}$ and the time integrals can be done following the procedure  discussed in Appendix \ref{app3}.


\section{Integration over $\{x_i\}$ and $\{\eta_i\}$}
\label{app3}

This appendix shows how to compute the integrals over  $\{x_i\}$ and $\{\eta_i\}$ appearing in the $R^{\left\lbrace\Phi,\Psi\right\rbrace}_{l}$ coefficients in \eqref{regphi}. These terms have the general form
\begin{eqnarray}
\overbrace{\vphantom{\int_0^1\frac{\text{d}x_1}{\sqrt{x_1}}}\int\text{d}\eta_1\cdots\int\text{d}\eta_{2N}}^{2N}\,\overbrace{\int_0^1\frac{\text{d}x_1}{\sqrt{x_1}}\cdots\int_0^1\frac{\text{d}x_{2N+1}}{{\sqrt{x_{2N+1}}}}}^{2N+1}\,\delta\left({\sum_{k=1}^{2N+1}x_k-1}\right)\,\left\lbrace \text{Pol}_1(\left\lbrace x_i\right\rbrace,\left\lbrace \eta_i\right\rbrace)+\log\left[\sum_{k=1}^{2N+1}x_k\,a^2(\eta_k)\right]\, \text{Pol}_2(\left\lbrace x_i\right\rbrace,\left\lbrace \eta_i\right\rbrace)\right\rbrace\nonumber\\
\label{logpol}
\end{eqnarray}
where the logarithmic contribution is included in  
the $R^{\left\lbrace\Phi,\Psi\right\rbrace}_{l,\,\text{log}}$ part
of \eqref{rparts}, whereas the pure polynomic one coming from $\text{Pol}_1$ is included in $R^{\left\lbrace\Phi,\Psi\right\rbrace}_{l,\,\text{pol}}$. Notice that we have redefined $2l$ appearing in expression \eqref{regphi}, namely the power of $p$, to be $2N$ in \eqref{logpol} in order to highlight its importance in the following discussion. Since the polynomials only introduce trivial modifications of the following formulas, let us focus on the expression
\begin{eqnarray}
\overbrace{\vphantom{\int_0^1\frac{\text{d}x_1}{\sqrt{x_1}}}\int\text{d}\eta_1\cdots\int\text{d}\eta_{2N}}^{2N}\,\overbrace{\int_0^1\frac{\text{d}x_1}{\sqrt{x_1}}\cdots\int_0^1\frac{\text{d}x_{2N+1}}{{\sqrt{x_{2N+1}}}}}^{2N+1}\,\delta\left({\sum_{k=1}^{2N+1}x_k-1}\right)\,\log\left[\sum_{k=1}^{2N+1}x_k\,a^2(\eta_k)\right].
\label{log}
\end{eqnarray}
There are $2N+1$ variables $x_i$ from the Feynman trick and all of them are integrated from 0 to 1. There are also $2N+1$ time variables $\eta_i$, but only $2N$ of them are integrated. In 
particular, $\eta_{2N+1}$ is not integrated. In order to recover the expressions given in the text, we have renamed $\eta$ as $\eta_{2N+1}$, $\eta'$ as $\eta_{2N}$ and $\eta''$ as $\eta_{2N-1}$. From the general expression \eqref{log}, it is straightforward to prove that for  $a(\eta)=1$, the logarithm vanishes since $\sum_{k=1}^{2N+1}x_k=1$. Therefore, $R^{\left\lbrace\Phi,\Psi\right\rbrace}_{l,\,\text{log}}=0$ in  nonexpanding spacetimes.

First, we deal with the integration over the $\{x_i\}$. Defining new variables $y_i^2=x_i$ for $i=1,\cdots,2N+1$, this integration can be written over the $2N$ sphere
\begin{eqnarray}
\int_0^1\frac{\text{d}x_1}{\sqrt{x_1}}\cdots\int_0^1\frac{\text{d}x_{2N+1}}{{\sqrt{x_{2N+1}}}}\,\delta\left({\sum_{k=1}^{2N+1}x_k-1}\right)\,=\,2^{2N}\,\int_{S^{2N}}\text{d}^{2N}{\Omega}\,.
\end{eqnarray}
Then, the logarithm can be expressed as
\begin{eqnarray}
\log\left[\sum_{k=1}^{2N+1}y^2_k\,a^2(\eta_k)\right]\,=\,\log\left[a^2(\eta_{2N+1})\right]+\log\left[1+\sum_{k=1}^{2N}y^2_k\,\left(\frac{a^2(\eta_k)}{a^2(\eta_{2N+1})}-1\right)\right],
\end{eqnarray}
where we have used that $y^2_{2N+1}=1-\sum_{k=1}^{2N}y_k^2$. The first logarithm on the right-hand side is the usual logarithm of the scale factor which appears in dimensional regularization in a FRW metric and it cancels out at the end. On the other hand, since $\eta_{2N+1}$ is an upper limit in all the time integrations (see next subsection), we have $\eta_k\leq\eta_{2N+1}$ for $k=1,\cdots,2N$. Thus, considering expanding universes, the argument of the logarithm is of the form $1+x$ with $-1<x\leq1$. Hence, it can be Taylor expanded as
\begin{eqnarray}
\log\left[1+\sum_{k=1}^{2N}y^2_k\,\left(\frac{a^2(\eta_k)}{a^2(\eta_{2N+1})}-1\right)\right]\,=\,\sum_{j=1}^{\infty}\,\frac{(-1)^{j+1}}{j}\,\left[\sum_{k=1}^{2N}y^2_k\,\left(\frac{a^2(\eta_k)}{a^2(\eta_{2N+1})}-1\right)\right]^j,
\end{eqnarray}
where the last factor on the right-hand side can also be expanded using the multinomial theorem
\begin{eqnarray}
\left[\sum_{k=1}^{2N}y^2_k\,\left(\frac{a^2(\eta_k)}{a^2(\eta_{2N+1})}-1\right)\right]^j\,&=&\,\sum^j_{\substack{l_1,l_2,\cdots, l_{2N}=0\\\sum_{i=1}^{2N}l_i=j}}\frac{j!}{l_1!\,l_2!\cdots l_{2N}!}\,\prod_{m=1}^{2N} \left[y^2_m\left(\frac{a^2(\eta_{m})}{a^2(\eta_{2N+1})}-1\right)\right]^{l_{m}}.
\end{eqnarray}
Therefore, the integration over the $2N-$sphere reduces to an integration of this kind:
\begin{eqnarray}
\int_{S^{2N}}\text{d}^{2N}{\Omega}\,y_1^{2l_1}\,y_2^{2l_2}\cdots y_{2N}^{2l_{2N}}\,=\,\frac{\sqrt{\pi}\,\prod_{i=1}^{2N}\Gamma(\frac{1}{2}+l_i)}{2^{2N}\,\Gamma(N+\frac{1}{2}+\sum_{i=1}^{2N}l_i)}\equiv \frac{1}{2^{2N}}\Gamma\left[\lbrace l_i\rbrace, 2N\right].
\end{eqnarray}
Then,
\begin{eqnarray}
2^{2N}\int_{S^{2N}}\text{d}^{2N}{\Omega}\,\log\left[1+\sum_{k=1}^{2N}y^2_k\,\left(\frac{a^2(\eta_k)}{a^2(\eta_{2N+1})}-1\right)\right]\,=&&\,\\\sum_{j=1}^{\infty}\,\frac{(-1)^{j+1}}{j}\,\sum^j_{\substack{l_1,l_2,\cdots, l_{2N}=0\\\sum_{i=1}^{2N}l_i=j}}\frac{j!}{l_1!\,l_2!\cdots l_{2N}!}\,&\Gamma\left[\lbrace l_i\rbrace, 2N\right]&\prod_{m=1}^{2N} \left(\frac{a^2(\eta_{m})}{a^2(\eta_{2N+1})}-1\right)^{l_{m}}.\nonumber
\end{eqnarray}
Applying the binomial theorem to the last factors,
\begin{eqnarray}
\left(\frac{a^2(\eta_{m})}{a^2(\eta_{2N+1})}-1\right)^{l_{m}}\,=\,\sum_{i_m=0}^{l_m}\,(-1)^{l_m-i_m}\,{l_m \choose i_m}\, \left[\frac{a^2(\eta_m)}{a^2(\eta_{2N+1})}\right]^{i_m}\,,
\end{eqnarray}
and gathering all the results we get
\begin{eqnarray}
&2^{2N}&\int_{S^{2N}}\text{d}^{2N}{\Omega}\,\log\left[1+\sum_{k=1}^{2N}y^2_k\,\left(\frac{a^2(\eta_k)}{a^2(\eta_{2N+1})}-1\right)\right]\,=\,\\&&-\sum_{j=1}^{\infty}\ \sum^j_{\substack{l_1,l_2,\cdots, l_{2N}=0\\\sum_{i=1}^{2N}l_i=j}}\frac{(j-1)!}{l_1!\,l_2!\cdots l_{2N}!}\,\Gamma\left[\lbrace l_i\rbrace, 2N\right]\,\sum_{i_1,i_2,\cdots, i_{2N}=0}^{l_1,l_2,\cdots, l_{2N}}\,(-1)^{\sum_{m=1}^{2N}i_m} \prod_{m=1}^{2N} \,{l_m \choose i_m}\left[\frac{a^2(\eta_m)}{a^2(\eta_{2N+1})}\right]^{i_m}\nonumber\,.
\end{eqnarray}

Finally, the time integrations can be done in a straightforward way since the dependence on $\eta_m$ of the scale factor is polynomial for the cosmologies considered in this work.

\subsection*{$R^\Psi_{l,\,\text{log}}=0$ for all cosmologies}

In Sec.\ \ref{cases}, it is mentioned that the $R^\Psi_{l,\,\text{log}}$ coefficients are all zero for all the cases considered. In fact, these expressions vanish not because of the integration over ${\{x_i\}}$ but because the polynomial $\text{Pol}_2(\{x_i\},\{\eta_i\})$ in \eqref{logpol} is zero for the $\Psi$ contribution. This can be shown by summing the already regularized expression for \eqref{ppsi}. Although the limits of integration are apparently different in each of the terms \eqref{ppsi1}, \eqref{ppsi2}, the region of integration is the same. For instance, the first integral can be written as
\begin{eqnarray}
\int_0^\eta \text{d}\eta' \left(\prod_{i=1}^{2l-1}\int_{\eta'}^{\eta}\text{d}\eta_i\right)\,=\,\int_0^\eta \text{d}\eta' \left(\prod_{i=1}^{2l-1}\int_{0}^{\eta}\text{d}\eta_i\,\theta(\eta_i-\eta')\right)\,,
\end{eqnarray}
where $\theta$ is the step function, while %
\begin{eqnarray}
\int_0^\eta \text{d}\eta' \int_0^{\eta'} \text{d}\eta'' \left(\prod_{i=1}^{2l-2}\int_{\eta''}^{\eta}\text{d}\eta_i \right)\,=\,\int_0^\eta \text{d}\eta' \int_0^{\eta} \text{d}\eta'' \theta(\eta'-\eta'') \left(\prod_{i=1}^{2l-2}\int_{0}^{\eta}\text{d}\eta_i \,\theta(\eta_i-\eta'')\right).
\end{eqnarray} 
Then, redefining in the last integral $\eta'$ as $\eta_{2l-1}$ and $\eta''$ as $\eta'$, both integrals have the same form
\begin{eqnarray}
\overbrace{\int^\eta_0\text{d}\eta_1\cdots \int^\eta_0\text{d}\eta_{2N}}^{2N}\,\prod_{i=1}^{2N-1}\theta(\eta_i-\eta_{2N})\,.
\end{eqnarray}
%


\twocolumngrid

\end{document}